\documentclass[twocolumn,pre,aps,showpacs,amsmath,amssymb]{revtex4-1}
\usepackage{graphicx}
\usepackage{bm}
\usepackage{epstopdf}
\begin{document}

\title{Work distribution in thermal processes}

\author{Domingos S. P. Salazar}
\email[]{salazar.domingos@gmail.com}
\affiliation{Unidade Acad\^{e}mica de Educac\~{a}o a Dist\^{a}ncia e Tecnologia, Universidade Federal Rural de Pernambuco, Recife, Pernambuco 52171-900 Brazil}

%\date{\today}

\begin{abstract}
We find the moment generating function (mgf) of the nonequilibrium work for open systems undergoing a thermal process, ie, when the stochastic dynamics maps thermal states into time dependent thermal states. The mgf is given in terms of a temperature-like scalar satisfying a first order ODE. We apply the result to some paradigmatic situations: a levitated nanoparticle in a breathing optical trap, a brownian particle in a box with a moving piston and a two state system driven by an external field, where the work mgfs are obtained for different timescales and compared with Monte Carlo simulations.
\end{abstract}
\maketitle

%\section{Introduction}
\textit{Introduction --} 
In thermodynamics, work is usually seen as a deterministic quantity of the process driving the system. In small systems far from equilibrium, as fluctuations turn relevant, its random behavior becomes apparent. In those irreversible cases, work depends not only on the driving process (or protocol), but also on the system's trajectory in the phase space. Resetting the system and repeating the same protocol will likely result in different values for work, $W_t$. In order to encode such randomness, the nonequilibrium probability distribution (pdf), $P(W_t)$, plays an important role in stochastic thermodynamics \cite{Seifert2012,Bustamante2005,Esposito2009,Sekimoto2010,Harris2007}. It may be represented in terms of its moment generating function (mgf):
\begin{equation}
\label{mgfdef}
    G(s,t) \equiv \langle e^{sW}\rangle = \int e^{sW}P(W_t=W)dW.
\end{equation}

Knowledge of the mgf allows the computation of statistical moments used in the optimization
of thermal engines \cite{Manikandan2019,Dechant2015,Verley2014}. In this context, a peculiar behavior of the work mgf is given by the Jarzynski equality (JE) \cite{Jarzinski1997}, $G(-\beta,t)=e^{-\beta \Delta F}$, where $\Delta F$ is the variation of free energy in the process, starting from thermal equilibrium. Together with other fluctuation theorems (FTs) \cite{Crooks1998,Jar2004,Campisi2009,Cuetara2014}, they found a broad range of applications in classic and quantum systems \cite{Wang2002,Liphardt2002,Collin2005,Alemany2012,Zhang2015,Hoang2018,Parrondo2015,Timparano2019,Hasegawa2019}. 

Going beyond the FTs seems to require more information about the specific system, which explains a noticeable lack of additional general results on the work mgf \cite{Seifert2012}. The goal of finding such insights is enriched by recent discussions about proper definitions of work \cite{TalknerPRE2017,Suomela2014,Elouard2017,LlobletPRL2017,Allahverdyan2014,Miller2017,Funo2018}, particularly in terms of quantum jump processes \cite{Hekking2013,Solinas2015,Suomela2016}, which allows the use of ideas from counting statistics and large deviation theory \cite{Garrahan2010}. In this framework, analysis is mostly focused on the long time behavior of the counting process, where the underlying mgf is related to the large deviations of the pdf. However, for finite time, universal features of the work mgf (\ref{mgfdef}) beyond the FTs are not clear, even for the simplest class of stochastic systems.

In this letter, we advance this subject by finding the work mgf (\ref{mgfdef}) for a class of stochastic with some minimal properties in their dynamics and thermal coupling. We are interested in a class that satisfies two properties: (i) the system maps thermal states into thermal states: starting at a thermal state with temperature $T_0$, after $t>0$, the system has a temperature governed by a time dependent law of cooling, $dT/dt=\phi_t(T)$, that contains information about the reservoir and the time dependent protocol. Additionally, (ii) the increment of work is proportional to the system's stochastic energy, $dW = \alpha(t) E(t) dt$, for any time dependent protocol function $\alpha(t)$ and stochastic energy $E(t)$ (continuous or discrete spectrum). In this case, we show that
\begin{equation}
\label{mgf}
       \log G(s,t) = s \int_0^t \alpha(u) U(\theta(u)) du, 
\end{equation}
with $U(\theta)=-\partial_\beta \log Z(\beta)$ as the (equilibrium) internal energy, $\beta=\theta^{-1}$ ($k_B=1$), and $\theta(t)$ is a temperature-like scalar that solves the following ODE
\begin{equation}
\label{ode}
    \dot{\theta}=\phi_t (\theta) + s\alpha_t \theta^2,
\end{equation}
with initial condition $\theta(0)=T_0$. Intuitively, properties (i) and (ii) are commonly found in simple systems with noninteracting degrees of freedom. It is the case of classic particles in a box \cite{Crooks2007,GONG16}, and also limiting cases of the Langevin dynamics \cite{Kwon2013}, such as a levitated nanoparticle in high vacuum \cite{Gieseler2012,Gieseler2018,Aspelmeyer2014}, as well as the usual overdamped limit of particles in liquid \cite{Speck2011,Kwon2013}. For system with a discrete spectrum, properties (i) and (ii) appear in particular Markov approximations, for instance, for a driven two level system \cite{Verley2013} and a one-step linear Markov process \cite{VanKampen}. All mentioned systems satisfy (\ref{mgf}), as discussed below.

The letter is organized as follows: First, we set the formalism and prove relation (\ref{mgf}). Then, we apply the result in some situations: a classic levitated particle in a single well potential, highlighting the harmonic and particle in a box as limiting cases, and a modulated two level system. For all applications, we compare the theoretical mgf with Monte Carlo simulations with excellent agreement.

%\section{Demonstration}
\begin{figure}[ht]
\includegraphics[width=3.3 in]{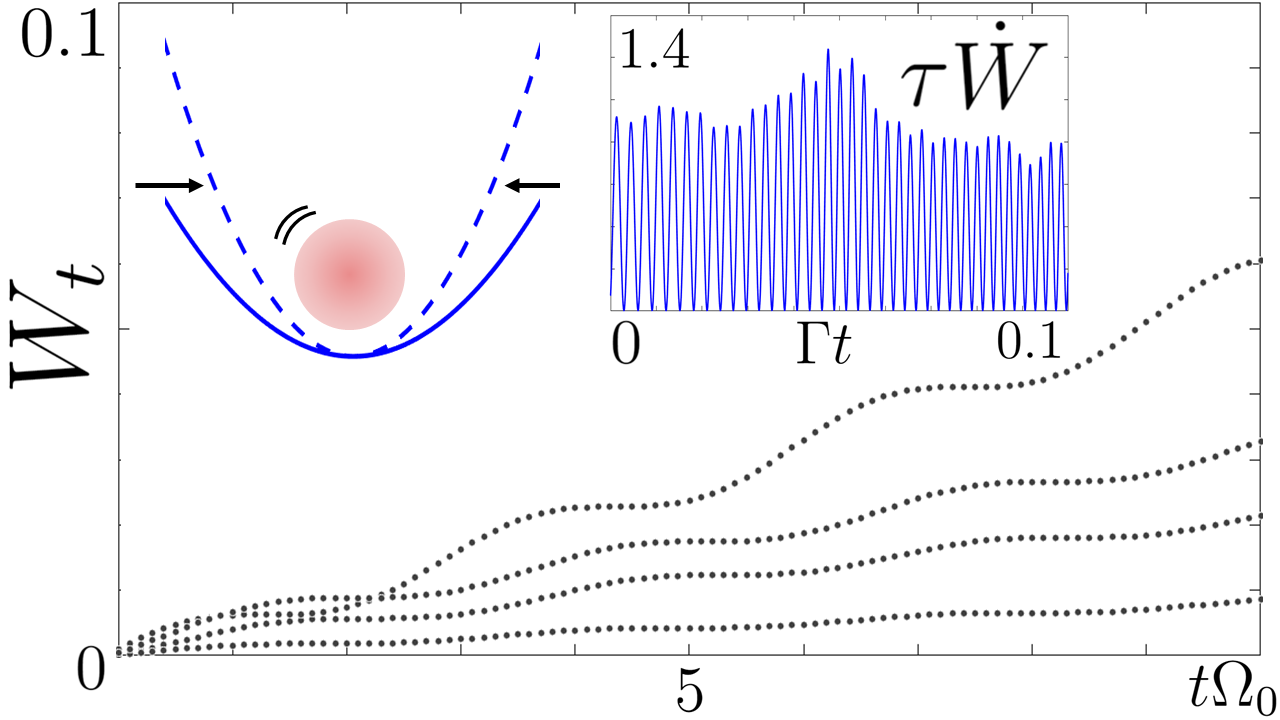}
\caption{The frequency of the optical trap changes in time, generating work over a levitated nanoparticle in high vaccum ($\Omega_0/\Gamma=10^3\gg1$). Sample trajectories of the stochastic work (in units of $k_B T$) as function of time for a particle initially prepared in equilibrium, subjected to a compression protocol $V_t^{-2}=\Omega_t^2=2^{t/\tau}$, with total duration $\Gamma\tau=0.1$, simulated with the Langevin equation (\ref{Langevin}). In larger timescales, the random fluctuations in the work increments, $\dot{W}=(W(t+dt)-W(t))/dt$, due to thermal coupling become apparent (see inset).}
\label{fig1}
\end{figure}
\textit{Formalism--} We prove the main result (\ref{mgf}-\ref{ode}). Formally, we assume (i) the open dynamics propagates thermal states into thermal states: given a thermal state $p(E|T)=Z(\beta)^{-1}g(E,\beta) e^{-\beta E}$, where
we omitted the dependency of $Z(\beta)$ on the protocol (as the final result remains unchanged) let $g(E,\beta)$ be the density of states, and transition probability $(E_0,t_0)\rightarrow(E_1,t_1)$ denoted by $R(E_0,t_0; E_1,t_1)$. We have
\begin{equation}
\label{propagator}
    \int dE_1 p(E_0|T_0) R(E_0,t_0; E_1, t_1) = p(E_1|T(t_1)),
\end{equation}
where $T(t)$ is the solution of a general law of cooling, $\dot{T}=\phi_t(T)$, with initial condition $T(t_0)=T_0$. Additionally, we assume (ii) the stochastic work satisfies
\begin{equation}
\label{workdef}
    d W_t = \alpha(t) E(t)dt,
\end{equation}
where $E(t)$ is the time dependent energy random variable (continuous or discrete spectrum) and $\alpha_t$ is some controllable protocol.
Our goal is to use properties (\ref{propagator}) and (\ref{workdef}) to find the work mgf, given the initial distribution, $p(E_0|\theta_0)$, is thermal. We split the time interval into $N+1$ discrete steps of length $\epsilon=t/(N+1)$, such that $W_t=\sum_{0}^{N} w_i$, with $w_i=\alpha_i \epsilon E_i$. Therefore, the work mgf is written as
\begin{equation}
\label{mgfsplit}
   G(s,t) = \int \prod_{i=0}^{N} dE_i   R_{i+1}^{i} p(E_0|\theta_0) e^{s \sum_0^{N} w_i},
\end{equation}
where $R_{i+1}^{i}=R(E_i,t_i; E_{i+1},t_{i+1})$ is a shorthand notation for the propagator from $t_i$ to $t_{i+1}=t_i+\epsilon$. First, notice that the exponential factor in $w_0=\alpha_0 \epsilon E_0$ may be combined with the thermal distribution as follows:
\begin{figure}[ht]
\includegraphics[width=3.3 in]{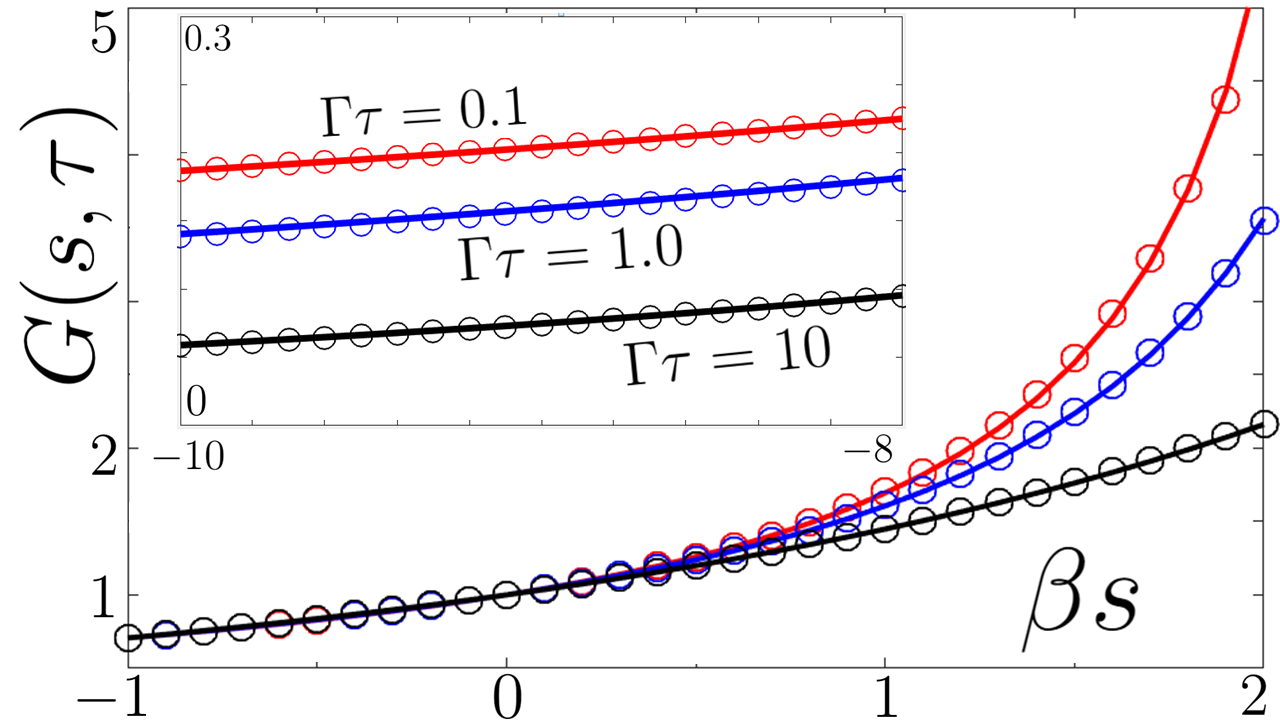}
\caption{(Color online) Theoretical work mgf in solid lines for the exponential protocol $V_t^{-2}=\Omega_t^2=\lambda_0 2^{t/\tau}$ vs. Monte Carlo simulations (symbols) of the Langevin equation (\ref{Langevin}) in the highly underdamped limit ($\Omega_0/\Gamma = 10^3$), using $m=10^6$ sample paths of the nanoparticle. Three protocols duration are showed: $\Gamma \tau= 0.1$ (red), $\Gamma \tau = 1$ (blue), $\Gamma \tau = 10$ (black). Notice the mgfs agree at $G(-\beta,\tau)=e^{-\beta \Delta F}=\sqrt{1/2}$ (JE) and $G(0,\tau)=1$, but they differ otherwise (see inset). Blow up values $G(s,\tau)=\infty$ for some $s>s^*(\tau)$ are also predicted by the theory.}
\label{fig2}
\end{figure}
\begin{equation}
\label{step1}
    p(E_0|\theta_0)e^{s\alpha_0 \epsilon  E_0}=\frac{Z(\beta_0-s \alpha_0 \epsilon )}{Z(\beta_0)}p(E_0|\theta_0'),
\end{equation}
with $\theta_0^{-1}=\beta_0$, yielding another thermal distribution with temperature $\theta_0'=(\beta_0-s\alpha_0 \epsilon)^{-1}\approx \theta_0(1+s\theta_0 \alpha_0 \epsilon)+\mathcal{O}\epsilon^2$. 
Second, from the thermalization property (\ref{propagator}), we solve the integral in $E_0$ and obtain
\begin{equation}
\label{step2}
    \int dE_0 p(E_0|\theta_0') R_1^0=p(E_1|\theta_0'+\phi_0(\theta_0')\epsilon)=p(E_1|\theta_1),
\end{equation}
for $\theta_1=
\theta_0'+\phi_0(\theta_0')\epsilon=\theta_0+ \phi(\theta_0)\epsilon + s \alpha_0 \theta_0^2 \epsilon +\mathcal{O}\epsilon^2$. Finally, using (\ref{step1}) and (\ref{step2}) in (\ref{mgfsplit}) results in
\begin{equation}
\label{mgfsplit2}
   G(s,t) = \frac{Z(\beta_0-s \alpha_0 \epsilon )}{Z(\beta_0)}\int \prod_{i=1}^{N} dE_i
  R_{i+1}^i  p(E_1|\theta_1) e^{s\sum_1^{N} w_i } .
\end{equation}
Comparing expressions (\ref{mgfsplit2}) and (\ref{mgfsplit}), we see the remaining integral above is a work mgf (\ref{mgfsplit}) computed for initial temperature $\theta_1$ and a protocol $\alpha(t)$ in the interval $(\epsilon,t)$. Repeating steps (\ref{mgfsplit})-(\ref{step2}) in (\ref{mgfsplit2}), one obtains by induction:
\begin{equation}
\label{logmgf}
    \log G(s,t) =  \sum_{i=0}^{N}
    \log \frac{Z(\beta_{i}-s\alpha_i \epsilon)}{Z(\beta_{i})}\approx s \sum_{i=0}^N \alpha_i U(\theta_{i})\epsilon +\mathcal{O}\epsilon^2,
\end{equation}
where $\beta_i^{-1}=\theta_i$. The scalar $\theta_i$ satisfies the map
\begin{equation}
\label{discrete}
    \theta_{i+1}=\theta_i + (\phi(\theta_i)+s \alpha_i \theta_i^2)\epsilon + \mathcal{O}\epsilon^2.
\end{equation}
Taking the limit $\epsilon \rightarrow 0$ in (\ref{logmgf}) and (\ref{discrete}) results exactly in (\ref{mgf}) and (\ref{ode}), respectively. In the next sections we apply the method to different examples. In each case, we show the system satisfies (i) and (ii). Then, we use equations (\ref{mgf}-\ref{ode}) to find mgfs and compare with Monte Carlo simulations.

%\section{Application: Levitated nanoparticles}
\begin{figure}[ht]
\includegraphics[width=3.3 in]{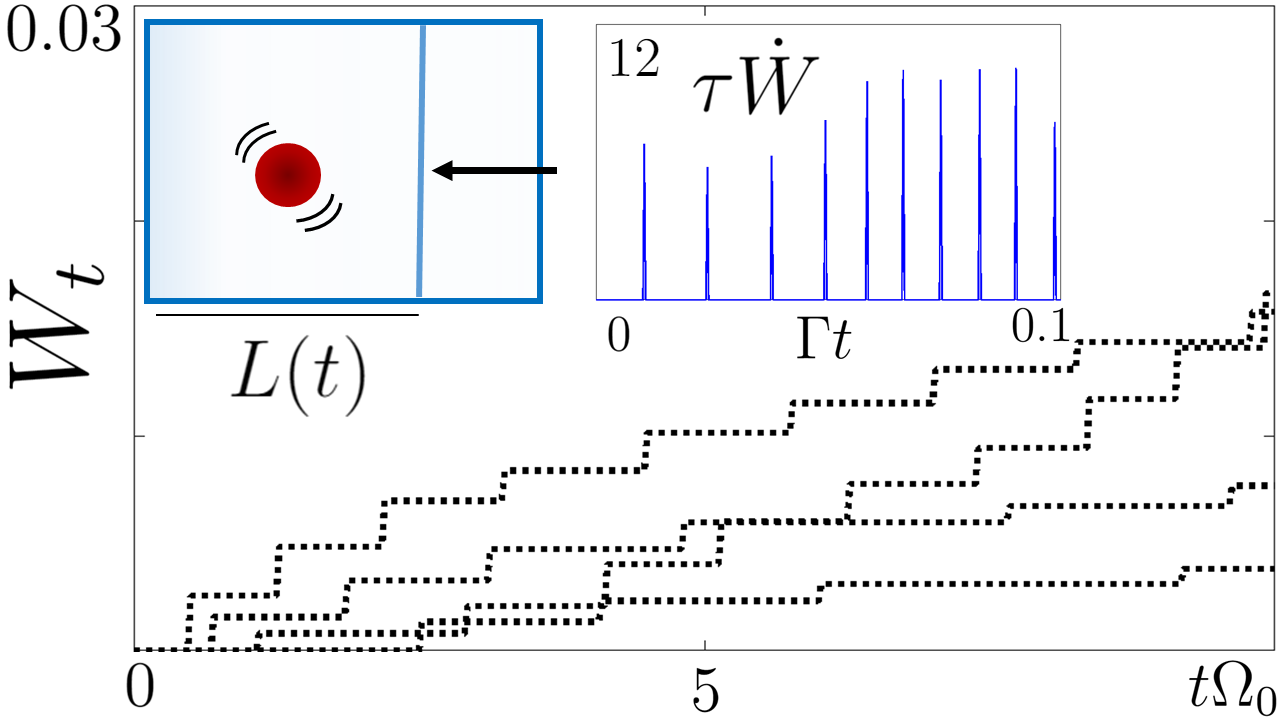}
\caption{The piston moves in time, $L(t)$, generating work over the box whenever it collides with the particle. Four sample trajectories of the stochastic work (units of $k_B T$) are displayed with stepwise behavior due to sparse collisions, for a single particle in a one dimensional box initially prepared in equilibrium with temperature $T$ (and $\sqrt{k_B T}/2L_0\equiv \Omega_0\gg\Gamma$), subjected to a compression protocol $L(t)=L_0 2^{-t/4\tau}$, and total protocol duration $\Gamma\tau=1.0$. In a larger timescales, the random fluctuations become apparent in the work increments, $\dot{W}=(W(t+dt)-W(t))/dt,$ due to thermal coupling (see inset).}
\label{fig3}
\end{figure}
\textit{Single well potential --}
For a representative open classic system, we consider the Langevin dynamics with a family of potentials $\mathcal{U}(x,k_t)=m k_t x^{2n}/2n$, and $n\geq 1$, which comprises the harmonic and box cases, as discussed bellow. The time-dependent stiffness is controlled by tuning parameter $k_t$. The particle's dynamics is given by
\begin{equation}
    \label{Langevin}
    \ddot{x} + \Gamma \dot{x} +\Omega_t^2 \ell_t \big(\frac{x}{\ell_t}\big)^{2n-1} = \frac{1}{m}F_{fluc}(t),
\end{equation}
for position $x(t)$, with gaussian noise $\langle F_{fluc}(t)F_{fluc}(t')\rangle = 2m\Gamma T \delta(t-t')$, where $\Gamma$ is a friction coefficient, $m=1$ is the particle mass, $T$ is the reservoir temperature and $k_t=\Omega_t^2 \ell_t^{2-2n}$. Define a characteristic volume $V_t=(T k_t^{-1})^{1/2n}=T^{1/2n}\Omega_t^{-1/n} \ell_t^{(n-1)/(n+1)}$ and the system's total energy, $E(x,p)=p^2/2+\mathcal{U}(x,k_t)$, with momentum $p=\dot{x}$, the following SDE is obtained for the energy in the highly underdamped limit \cite{Gieseler2012}, $\sqrt{T}/V_t\gg\Gamma$:
\begin{equation}
\label{LangevinforE2}
dE=-\Gamma_n (E-\frac{f_n}{2} T)dt+\sqrt{2\Gamma_n TE}d\textrm{B}_t -\frac{2\dot{V_t}}{f_n V_t}E,
\end{equation}
with $\Gamma_n/\Gamma=2n/(n+1)$, and fractional degrees of freedom $f_n=(n+1)/n$ \cite{Salazar2019A},  $d\textrm{B}_t$ is a Wiener increment. Notably, the dynamics (\ref{LangevinforE2}) maps thermal states into thermal states \cite{Salazar2019A}, satisfying property (i). 
Moreover, the last term represents work, $dW = \dot{k} \partial_k \mathcal{U}(x,k) \approx - 2\dot{V}/(f_n V)E$, where the approximation follows from the virial theorem (provided $\dot{V}/V$ changes slowly over an oscillation) which satisfies property (ii). Taking the ensemble average of (\ref{LangevinforE2}) and using $\langle E \rangle=U(T)=(f_n/2)T$, the law of cooling reads
\begin{figure}[ht]
\includegraphics[width=3.3 in]{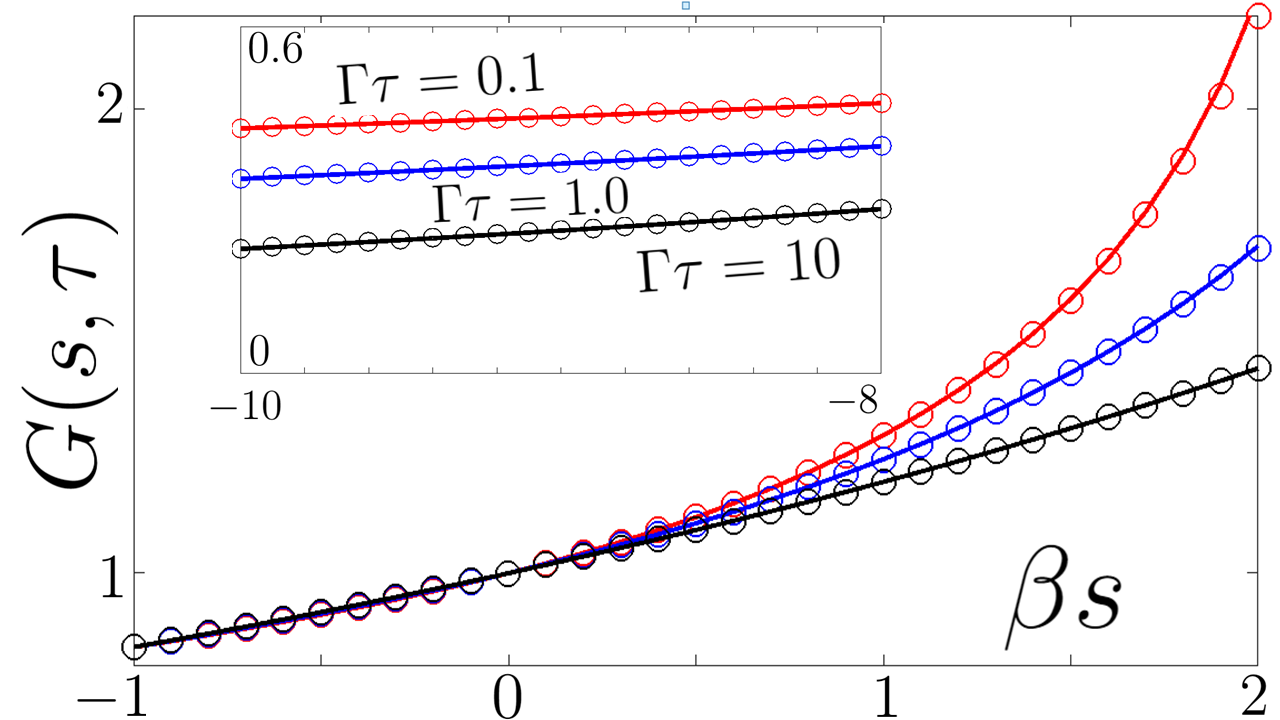}
\caption{(Color online) The theoretical work mgf $G(s,\tau)$ of the brownian particle in a box (\ref{mgfexpo}) in solid lines for the exponential protocol $L(t)=L_0 2^{-t/4\tau}$ vs. Monte Carlo simulation of the brownian particle (symbols) for a slow piston ($\sqrt{k_b T}/(2L_0)\approx 10^2\Gamma$). For the numerical mgf, we used $m=10^6$ samples of the single particle. Three protocol durations are showed: $\Gamma \tau= 0.1$ (red), $\Gamma \tau = 1$ (blue), $\Gamma \tau = 10$ (black). Notice the mgfs agree at $G(-\beta,\tau)=e^{-\beta \Delta F}=\sqrt{1/2}$ (JE) and $G(0,\tau)=1$, but they differ otherwise (see inset).}
\label{fig4}
\end{figure}
$\phi_t(\theta)=-\Gamma_n(\theta-T)+\alpha_t \theta$, with $\alpha_t=-2\dot{V}/(f_n V)$, which replacing in (\ref{ode}) yields the following scalar Ricatti equation
\begin{equation}
\label{Ricatti}
    \dot{\theta}=-\Gamma_n(\theta-T) - \theta (1+s\theta) \frac{2\dot{V_t}}{f_n V_t}.
\end{equation}
Finally, the work mgf (\ref{mgf}) is given by the solution of (\ref{Ricatti}). As an illustration, we show how the exponential protocol ($\dot{V}/V=const$) yields a closed form mgf.

\textit{Exponential protocol--}
Consider the dynamics (\ref{Langevin}) under a exponential protocol $V(t)=V_0 e^{-f_n B t/2}$ with constant $B$ and prepared in equilibrium with the reservoir (temperature $T$). The protocol presents an interplay between heat (dissipation) and work in the stochastic dynamics (Fig. 1). Such protocol appears as optimal solutions to the minimum entropy production problem
\cite{GONG16,Dechant2017}. In this case, the scalar Ricatti equation (\ref{Ricatti}) with $-2\dot{V}/(f_n V)=B$ has a solution
\begin{equation}
\label{solutionexpo}
   \theta(t)=\frac{1}{sB}[
   \frac{\sqrt{D(s)}}{2}\tan(\frac{\sqrt{D(s)}}{2}(t+c_1))+\frac{\Gamma_n-B}{2}],
\end{equation}
with constant $c_1$ from initial condition $\theta(0)=T$ and $D(s) = -(B-\Gamma_n)^2+4B\Gamma_n s T$. Finally, inserting (\ref{solutionexpo}) in (\ref{mgf}) with $U(\theta)=(f_n/2)\theta$ results in
\begin{equation}
\label{mgfexpo}
    G(s,t)=\frac{e^{f_n(\Gamma_n-B)t/4}
}{(\cosh(\omega(s)Bt)+\sinh(\omega(s)Bt)g(s))^{f_n/2}},
\end{equation}
with $g(s)= -i(D(s) + B^2-\Gamma_n^2)/(2\Gamma_n\sqrt{D(s)})$ and $i \omega(s)=\sqrt{D(s)}/2B$. Notice that $\beta \Delta F=(f_n/2)Bt$,
%the mgf can also be rewritten as
%\begin{equation}
%\label{mgfexpo2}
%    G(s,t) = \frac{e^{-\beta \Delta F(1+ \kappa(s)-\kappa(-\beta))}}{1+h(s)(1-e^{-2\kappa(s) \beta \Delta F})},
%\end{equation}
%with $\kappa_s = \sigma \omega(s)$ , $h(s)=(\sigma f(s)-1)/2$, $\sigma=sign(\Gamma+B)$. In this notation, notice $h(-\beta)=0$, 
which makes $G(-\beta,t)=e^{-\beta \Delta F}$ (JE) (see Sup. Mat). Also notice that there are values of $s^*$ such that $G(s^*,t)=\infty$, ie, the solution of the transcendental equation $\cosh(\omega(s^*)Bt)+\sinh(\omega(s^*)Bt)g(s^*)=0$. The particular sudden change case $\Gamma=0$ results in a known result \cite{Crooks2007} (see Sup. Mat.). Now we apply the mgf (\ref{mgfexpo}) for the relevant systems levitated nanoparticle in a laser trap ($n=1$) and a particle in a box ($n=\infty$).

\textit{Levitated nanoparticle ($n=1$)--}
A levitated nanoparticle in a optical trap satisfies (\ref{Langevin}) in the particular case $n=1$ \cite{Gieseler2012,Gieseler2018}. In the highly underdamped limit, attained with $\sqrt{T}V_t^{-1}=\Omega_t \gg \Gamma$, we have (\ref{LangevinforE2}) with $f_n=1$, $\Gamma_n=\Gamma$. In Fig. 1, we show some work samples for the exponential protocol $\Omega^2_t=2^{t/\tau}$. In Fig. 2, we compare the theory (\ref{mgfexpo}) with Monte Carlo simulations for the levitated nanoparticle using $m=10^6$ copies of the dynamics (\ref{Langevin}) (details in Sup. Mat.) for all ranges of protocol duration. For the divergence in $G(s^*,\tau)$ we obtain approximate values $s^*=\{2.46,2.91,8.23\}$ from theory for the cases $\tau \Gamma=\{0.1,1.0,10\}$, respectively.

\textit{Particle in a box ($n\rightarrow \infty$) --}
A brownian particle in a box is simulated with dynamics $\dot{v}+\Gamma v = F_{fluc}$ and reflecting walls. When the particle (velocity $v$) collides with the piston (velocity $v_p=dL/dt$) at $x=L(t)$, it reflects the particle elastically with velocity $2v_p-v$. Over the time interval $[t_0,t_1,...,t_{N+1}=\tau]$, this interaction produces work given by
\begin{equation}
\label{workbox}
W(\tau)=\sum_{i=0}^{N}2v_p(t_i)(v_p(t_i)-v(t_i))\delta(t_i),
\end{equation}
where $\delta(t_i)=1$ if there is a collision with the piston at $[t_{i},t_{i+1})$, and $\delta(t_i)=0$ otherwise \cite{GONG16}. In Fig. 3, we represent simulations of particle in a box for a exponential compression protocol $L(t)=L_0 2^{-t/4\tau}$. Differently from the harmonic case (Fig. 1), the work increments in the box are sparse (Fig. 3), and the total work behaves as a step-wise random function of time, one step for each collision with the piston. The box dynamics is also a particular case of the Langevin dynamics (\ref{Langevin}) for $n=\infty$, $\mathcal{U}=0$ for $|x|<V(t)=\ell_t$, $f_n=1$ and $\Gamma_n=2\Gamma$. In Fig. 4, we compare the theoretical work mgf (\ref{mgfexpo}) for the box with parameter $\dot{V}/V=\dot{\ell}/\ell=-1/4\tau$ ($B=1/\tau$), with Monte Carlo simulations (details in Sup. Mat.) for protocol $L(t)=L_0 2^{-t/4\tau}$ and compute $G(s,\tau)$ from (\ref{workbox}) and (\ref{mgfdef}) with different time duration ($\Gamma\tau={0.1,1.0,10}$) for a slow piston ($\sqrt{T}/2L_0 \approx 10^2 \Gamma \gg \Gamma$).

%For the simulations, we start a single particle at equilibrium and defined a natural frequency, $\sqrt{k_B T}/2L_0=\Omega_0$, choosing a small $L_0$ such that $\Omega_0/\Gamma \approx 10^2 \gg 1$ (see Sup. Mat.
%Then, we simulate the brownian motion ($dt=L_0/20$ for the timestep) with elastic walls and moving piston $L(t)$, using $n=10^6$ copies of the system and collecting the statistics $G(s,\tau)=\langle e^{sW_\tau} \rangle$. As in the case of levitated nanoparticles, the function $G(s,\tau)$ diverges for a finite value of $s=s^*$, yielding approximate values $s^*=\{2.XX,2.XX,8.XX\}$ for the cases $\tau \Gamma=\{0.1,1.0,10\}$, respectively.

\textit{Two level system--}
Consider a driven two level system coupled to a thermal bath with temperature $T$ initially prepared in equilibrium. The system has energy $E=-h_t\sigma$, with external field $h_t$ and $\sigma=\pm 1$. In term of jump processes \cite{Verley2013}, the work can be written as $\dot{W}=-\dot{h}\sigma=(\dot{h}/h)E$, and it naturally satisfies (ii). The transition rates are given by $\nu_{-\sigma,\sigma}=\nu(h)e^{-\beta \sigma h}$. Moreover, in this setup, a temperature may be defined for all times, which satisfies (i). We find the law of cooling (see Sup. Mat):
\begin{equation}
\label{2lvlcooling}
    \phi_t(\theta)=\frac{\dot{h_t}}{h_t}\theta+\frac{2\theta^2\nu(h_t)}{h_t}\cosh\big(\frac{ h_t}{\theta}\big)\sinh\big(h_t(\frac{1}{\theta}-\frac{1}{T})\big).
\end{equation}
\begin{figure}[ht]
\includegraphics[width=3.3 in]{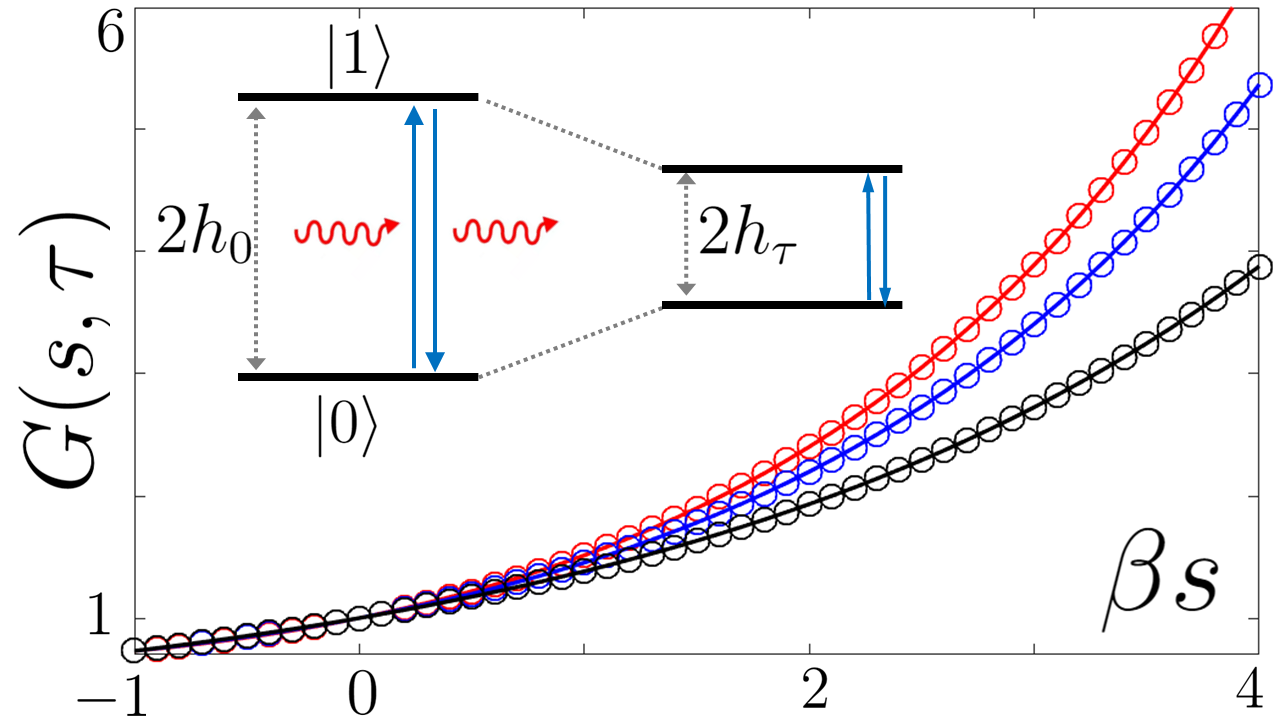}
\caption{(Color online) Theoretical work mgf $G(s,\tau)$ of the two level system in solid lines for the linear protocol $h(t)=h_0(1-t/2\tau)$ vs. Monte Carlo simulation (symbols), with $h_0/T=1$. For the simulation, we used $m=10^6$ samples of the system. Three protocol duration are showed: $\Gamma \tau= 0.1$ (red), $\Gamma \tau = 1$ (blue), $\Gamma \tau = 10$ (black). Work is computed as $dW=E(\dot{h}/h)dt$. Notice the three curves agree at $\beta s = 0$ and $\beta s = -1$ (JE).}
\label{fig5}
\end{figure}
Finally, applying the law of cooling (\ref{2lvlcooling}) in (\ref{ode}), we obtain $\dot{\theta}=\phi_t(\theta)+s(\dot{h}/h)\theta^2$. Then, computing (\ref{mgf}) with $U(\theta)=-h_t\tanh(h_t \theta^{-1})$, the work mgf is obtained. In Fig. 5, we show the work mgf for the protocol $h(t)=h_0(1-t/2\tau)$ integrated from (\ref{mgf}) numerically, such that $h(\tau)=h_0/2$, $T=1$ and Arrhenius rates $\nu(h)=\Gamma$, compared with Monte Carlo simulations (details in Sup. Mat.), also with excellent agreement for short and long protocol duration.

\textit{One-step linear process--} Consider a discrete energy system with energy $E(h,\sigma)=2h\sigma$, with states $\sigma=\{0,1,2,...\}$ and uniform energy gaps controlled by the protocol $h=h(t)$. Work is defined as $dW=\dot{h}\partial_h E(h,\sigma) dt = (\dot{h}/h)E(h,\sigma) dt$, satisfying property (ii). Take the dynamics of $p_n(t)=P(\sigma=n;t)$ defined as the one step linear process $\dot{p}_n(t)=r_{n+1}p_{n+1}+g_{n-1}p_{n-1}-(r_n+g_n)p_n$, with $r_n=\nu(h_t)an$ and $g_n=\nu(h_t)b(n+1)$, such that $b/a=\exp(-2h(t)/T)$ (local detailed balance) \cite{Harris2007}. For a constant $h_t$, this dynamics corresponds to the weakly coupled quantum harmonic oscillator \cite{VanKampen}. Starting from equilibrium, $p_n(0)=e^{-\beta h_0 n}/Z(\beta,h_0)$, we show (see Sup. Mat.) the dynamics satisfies property (i) with law of cooling
\begin{equation}
\label{QHOcooling}
    \phi_t(\theta)=\frac{\dot{h_t}}{h_t}\theta+\frac{2\theta^2\nu(h_t)}{h_t}\sinh\big(\frac{ h_t}{\theta}\big)\sinh\big(h_t(\frac{1}{\theta}-\frac{1}{T})\big),
\end{equation}
which is the driven version of the known thermal relaxation for a bosonic mode in the Lindblad's dynamics \cite{Denzler2018,Salazar2019B}. The mgf $G(s,t)$ is given by (\ref{mgf}) with $U(\theta)=2h(t)\exp(-2h(t)/\theta)/(1-\exp(- 2h(t)/\theta))$, where $\theta$ is the solution of (\ref{ode}) with $\phi_t (\theta)$ given by (\ref{QHOcooling}), and  $\alpha=\dot{h}/h$. Notice that the limit $(\beta h_t,\theta^{-1}h_t) \rightarrow 0$ (continuous spectrum) yields the Ricatti equation (\ref{Ricatti}) for the classic harmonic oscillator ($n=1$) in the highly underdamped limit, if $\nu(h_t)\rightarrow \Gamma/2\beta h_t$ (Bose rates).

\textit{Other applications--} The method is also suitable in the description of the overdamped limit ($\Omega_0/\Gamma \ll 1$) of the Langevin equation (\ref{Langevin}), for $n=1$, as it also satisfies (i) and (ii) \cite{Speck2011}. In this case, the system evolves thermally with the ODE (\ref{ode}) reading $\dot{\theta}=-2(k_t/\Gamma)(\theta-T)+(\dot{k}/k)\theta (1+s\theta)$ (see Sup. Mat). Which can be solved for $\theta_t$ and inserted in (\ref{mgf}). More generally, the result is also applicable to overdamped setups with controlled environment temperature \cite{Martines2015}, a situation in which the underlying ODE (\ref{ode}) will also depend on the temperature protocol.

\textit{Conclusions--}
We proposed a method to compute the work mgf for a class of stochastic thermal processes, with discrete or continuous energy spectrum. The nonequilibrium behavior of $G(s,t)$ is encoded in two ingredients: the internal energy $U(\theta)$ in (\ref{mgf}) and a temperature-like scalar, $\theta$, whose dynamics follows a modified law of cooling (\ref{ode}). The resulting mgf (\ref{mgf}) unifies some previous findings of seemly unrelated systems. We compared the theory with Monte Carlo simulations of levitated nanoparticles, a brownian particle in a box and a two level system, showing excellent agreement for different timescales. Other applications were briefly discussed. As future research directions, note thermal processes (i) are also found in special cases of gaussian channels with diagonal covariance matrices, for which the Lyapunov equation for the covariance matrix becomes a scalar law of cooling. More generally, a thermal process could also be enforced in quantum systems \cite{Alipour2019} such that (i) is feasible. However, a shortcoming of our approach is that property (ii) needs to be reformulated with a proper definition of work \cite{Funo2018} for quantum systems. Even considering the framework of quantum jumps \cite{Hekking2013}, a consistent quantum formulation of property (ii) is not clear due to the role of the measurement scheme. Such generalization is left for for further investigation.

\end{document}